# All-atom Molecular Dynamics Simulations of the Projection Domain of the Intrinsically Disordered htau40 Protein

N.E. Sanborn, N.R. Hayre, R.R.P. Singh, and D.L. Cox

Department of Physics and Institute for Complex Adaptive Matter, University of California, 1 Shields Ave., Davis, CA 95616 USA

**Abstract**

We have performed all atom molecular dynamics simulations on the projection domain of the intrinsically disordered htau40 protein. After generating a suitable ensemble of starting conformations at high temperatures, at room temperature in an adaptive box algorithm we have generated histograms for the radius of gyration, secondary structure time series, generated model small angle x-ray scattering intensities, and model chemical shift plots for comparison to nuclear magnetic resonance data for solvated and filamentous tau. Significantly, we find that the chemical shift spectrum is more consistent with filamentous tau than full length solution based tau. We have also carried out principle component analysis and find three basics groups: compact globules, tadpoles, and extended hinging structures. To validate the adaptive box and our force field choice, we have run limited simulations in a large conventional box with varying force fields and find that our essential results are unchanged. We also performed two simulations with the TIP4P-D water model, the effects of which depended on whether the initial configuration was compact or extended.

Corresponding author: N.E. Sanborn nehall@ucdavis.edu

## 1 Introduction

Tau is an intrinsically disordered protein (IDP) that plays a critical role in the axon of neurons, helping to stabilize and crosslink microtubules in bundles[1]. These bundles serve as the "highway" for anterograde and retrograde transport to and from the presynaptic terminal[2]. In the late stages of Alzheimer's disease and in chronic traumatic encephalopathy, the tau protein acquires extra phosphate groups and aggregates into hyperphosphorylated cross-beta structure "tangles" or filaments[3].

Clearly, the tau is an important protein for understanding both normally functioning neurons as well as how neurons can become nonfunctional in the aforementioned neurodegenerative diseases. Understanding the structure of even free tau monomers is difficult from an experimental standpoint because of the intrinsic disorder; high resolution structures of taus in microtubule bundles or tau filaments are beyond the scope of current diffraction or magnetic resonance approaches. Hence, there is a clear role for simulation and theory to help elucidate the properties and ensembles of structures for monomeric taus.

In particular, focusing studies on the so-called projection domain of the tau protein, which runs from the N-terminus to the putative microtubule binding domain, can help shed light on a number of important questions, including: How do these projection domains reach to lengths of about 20 nm [4] in cross-linking microtubules? Given that the microtubule binding domain is implicated in the cross beta-structure of the hyperphosphorylated tau filaments[5], is the projection domain relevant for understanding the disordered section of these filaments? In the microtubule bundles, what is the mechanical character of the taus and how does that change with the progression of disease? Finally, as a significant IDP, how does tau fit into the emerging understanding and classification of IDPs?

In this paper we use a variety of all-atom molecular dynamics simulations, including a novel adaptive box approach for large proteins, to generate conformation ensembles for the projection domain of the human htau40 protein, which we characterize by histograms of the radius of gyration and time series of secondary structure stability. This is significantly different than the work done by Battisti, who characterized full-length tau in the 100 picosecond range[6], as we have done 30 simulations of the tau projection domain (TPD) for 50 ns for an aggregate time of 1.5 μs, as well as 6 octahedral box simulations for 10 ns with a variety of force-fields and two additional conformations in an alternate water model. We compare our conformations to IDP theory and find that the tau fits within the Flory Random Coil regime of IDPs. From our numerical data we generate predictions for NMR chemical shifts and small angle x-ray scattering (SAXS) intensity which we compare to experiment. The SAXS profiles are similar to experimentally studied tau subsequences, although none of these overlap precisely with our studied sequence. We find that the chemical shifts compare most favorably to the disordered portion of tau filaments rather than full length free tau in solution. In addition, we categorize tau using principal component analysis (PCA) and radius of gyration in to determine the behavior over time, and to characterize whether our simulations are a good representation of tau in solution or another comparable system.

## 2  Methods

### 2.1 Generation of conformations

We simulated the first 242 residues, the so called projection domain, of human tau protein ht40, following the coordinate generation method of Staneva et al. [7]We generated starting structures in two different ways. First, we generated extended structures in Pymol[8] which were then simulated in implicit solvent for 20ns at T=300 using the AMBER suite program PMEMD on our GPU cluster [9, 10]. We used the ff12SB force-field for all coordinate generation runs. These simulations were done in order to place the structures into a more compact, energetically favorable state. During the simulation time, all Pymol generated structures collapsed into globules with approximately 3-4 nm radius of gyration. Several snapshots of varying length from two of these initial runs were then used in implicit solvent, high temperature runs (T=450K) to generate a wider sample of initial conformations. All five of these Pymol-generated conformations were then put into simulations with one end held down to approximate attachment to a microtubule. We also generated six initial structuresS using loopy, part of the jackal package[11], which generates a more random structure. Because the loopy structures are more compact and have energy considerations already taken into account, our initial runs for these structures were carried out at high temperature. These loopy-generated structures were simulated with the same conditions as the Pymol structures, but were not held down at one end. We selected a total of 20 snapshots from both Pymol- and loopy-generated high temperature runs with both collapsed and extended structures included for all atom simulations. An additional ten duplicate simulations were also performed with different random seeds.

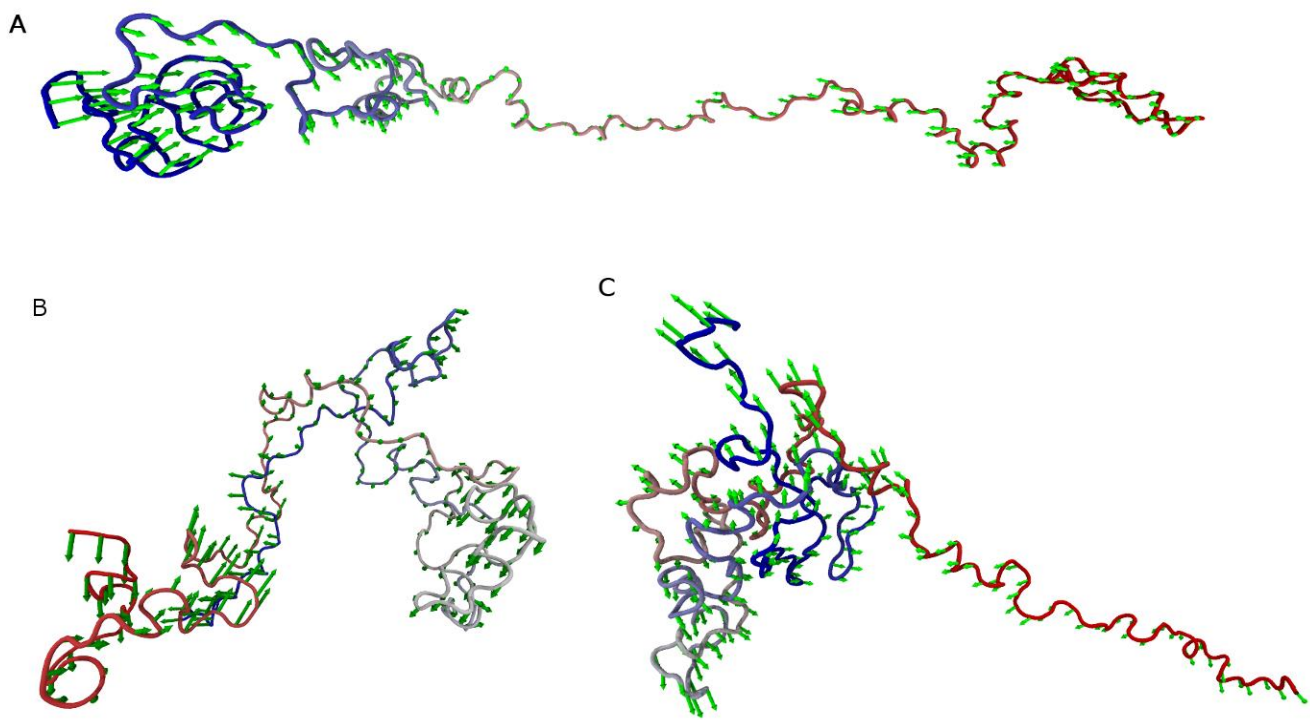

*Figure 1 PCA for extended (A), compact (B), and tadpole structures (C).* **Twenty of the thirty adaptive box runs were either compact or tadpole, and tended to get more compact over time. The ten extended runs both extended and compressed, but never became as compact as the other types of structures.**

*2.2 All atom simulations*

We performed thirty all atom simulations at T=300 K. These runs were carried out for 50 ns using an adaptive box method and the ff12SB force field and TIP3P water model, using the AMBER12 package on a GPU cluster.[9, 12] Instead of a large cubic or octahedral box, a small box with at least 1 nm padding at minimum surrounds the protein. When the protein moves too close to the edge of the box, the configuration of a shell of water molecules surrounding the protein is saved to preserve the hydrated conformation. A new box, refilled with water, is then formed around the protein with the same padding dimensions as the initial box. Under these conditions, explicit solvent runs of large proteins that take much less simulation time than using a cubic or octahedral box, but it makes free energy calculations more difficult since total waters molecules are not preserved.

Since the TIP3P water model is known to cause overly-compact IDP structures, we simulated two runs using the TIP4PD water model[13]. This force-field is a modification of the TIP4P water that is better at modeling IDPs than standard TIP4P or TIP3P[13]. We simulated both an extended structure and a more compact structure using the ff12SB force-field within an octahedral box. The extended structure further simulated with TIP3P water in five different force-fields: ff03.r1, ff03ua, ff10, ff99SB, and ff12SB. The compact structure was simulated in ff12SB only. The two ff12SB runs were then continued for an additional 10 ns. The purpose of these simulations was to assess the effect of the adaptive box and force fields. The large number of atoms in these structures and perimeter regions necessitated shorter simulation times. In addition, PCA, radius of gyration, and secondary structure analysis was also performed on the octahedral box runs. We did not analyze residual dipolar couplings, because they have been found to be inaccurate for simulations of IDPs even with IDP-specific force-fields[14]

*2.3 Analysis of trajectories*

Principal component analysis (PCA) was completed in using Normal Mode Wizard, a plugin included in Visual Molecular Dynamics (VMD)[15] that uses the ProDy package[16]. PCA was performed over complete runs to demonstrate the overall change in shape of the TPD. PCA was mainly completed for easier classification of runs over the 50ns simulation time, as it is not a reliable method for determining overall modes for such a large protein on this time scale[17]. We took snapshots from every 0.1 ns to input into Crysol. Crysol simulates SAXS data from the available PDB files[18]. Then, we averaged the individual

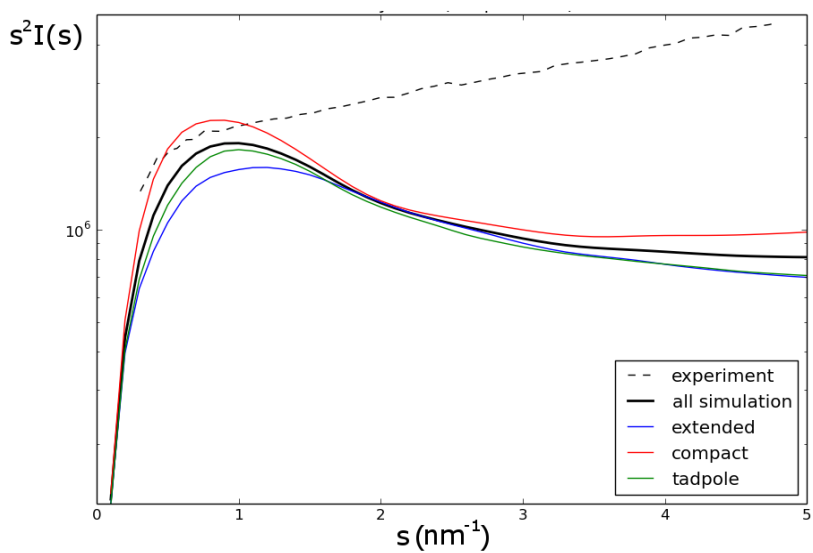

*Figure 2 Kratky plots compared with K23 experimental SAXS.* None of our simulated tau (solid) are completely random coils, but rather appear to be partially disordered. Interestingly, the extended structures are flatter and therefore closer to random coil than the compact or tadpole structures.

SAXS curves to achieve a look at the entire ensemble. This average curve was then compared to SAXS curves from experiment with similarly sized tau fragments, although an exact match to the ht40 TPD could not be found in the literature. Simulation of chemical shift information was extracted using shiftx2 [19] for each 50ns run. Shiftx2 creates simulated chemical shifts predicts chemical shifts for $^1$H, $^{13}$C, and $^{15}$N[19]. We focused on the $^{13}$C chemical shifts for comparison with experiment. To more easily compare the chemical shifts of our simulations with experiment, all chemical shift data was averaged. This allows for a broader comparison than looking at secondary structure alone, especially since the tau protein is unstructured and should have no consistent secondary structure. We also computed dihedral angles using VMD[15] to analyze secondary structure formation directly from the PDB file itself for further comparison.

### 2.4 Theoretical classification of Tau as an IDP

In order to compare with other types of random coils, $R_{ij}$ profiles were calculated[20] using the time-averaged structures created for SAXS analysis. $R_{ij}$ is the physical distance between residues $i,j$. The $R_{ij}$ profile is generated by plotting $R_{ij}$ vs. $|i - j|$, which is the sequence separation between the residues. This parameter describes how extended or compact the protein is and whether it forms a globule, a Flory random coil, or an extended structure. Finally, AMBERtools [18] was used to calculate radius of gyration for each frame of each run.

## 3 Results

### 3.1 Structure

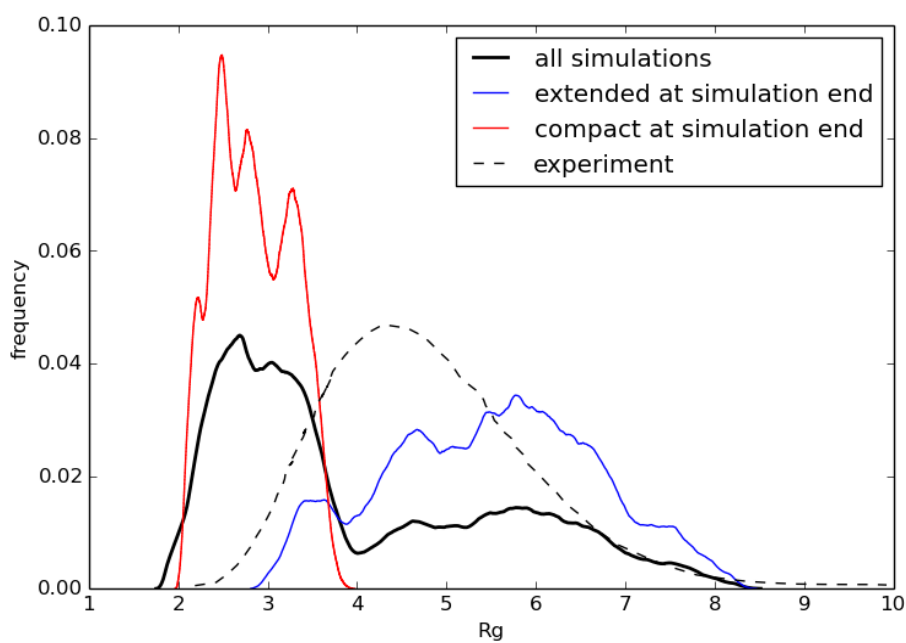

*Figure 3 Radius of gyration histograms of simulation compared with the K23 fragment [21].* All have been normalized so that the integral is equal to one. Note that our adaptive simulations effectively have two groupings, the more extended of which is closer to the experimental distribution, although peaked at a higher radius of gyration.

TPDs simulated in implicit solvent conditions all collapsed into globules within 20ns simulation time at room temperature. High temperature implicit solvent required more time until collapse, about 50ns, while clearly exploring additional conformations. Holding the binding domain end fixed during simulation also prolongs the time to collapse, but implicit solvent simulations still eventually reach a collapsed state with a radius of gyration of around 3-4 nm. The increase in time between free protein and one fixed end implicates that both charged ends move toward each other in the free TPD, while only one end can move towards the other when one end is held down. We believe the uniformity of collapse in implicit solvent is caused by a missing cutoff of electrostatic forces in GPU based AMBER and the fact that water is approximated as a static dielectric. This overemphasizes opposite charge attraction and enhances the tendency to collapse.

Fully solvated, all atom adaptive box simulations do not collapse even for runs as long as 100ns. PCA allows us to separate the runs into three rough categories: extended structures with no globule-like areas, compact globules, and tadpoles with the n-terminus end extended (fig. 1). The extended structures are not globule-like, and either tend toward opening or closing in an accordion-like or hinge-like fashion. Compact structures tend to have both ends of the tau fragment folded roughly in half, and flatten over time with respect to the long axis of the protein. Tadpoles are seen when the c-terminus end forms a globule, while the n-terminus end remains extended.

Experimental SAXS[21] shows that tau is disordered, while model SAXS data from our simulation appears to only be partially disordered, suggesting our simulations have more secondary structure. The model SAXS curve shows less extension than the experiments (fig. 2). The radius of gyration for the K23 fragment, which is 254 residues is 4.9 ±0.2nm. The average radius of gyration from the crysol generated SAXS curves is 3.8±0.3nm, which is similar to the average calculated directly from the PDB files of the simulation results (4.0±0.3nm). Tau in solution is therefore more extended than our adaptive box model of tau, as well as more disordered. The radii of gyration can be separated into two distinct groups for our simulation (fig. 3), one of which is slightly more extended than K23, and the other of which is are compact and globule-like. The overall shape of the extended structures histogram is closer to that of experiment.

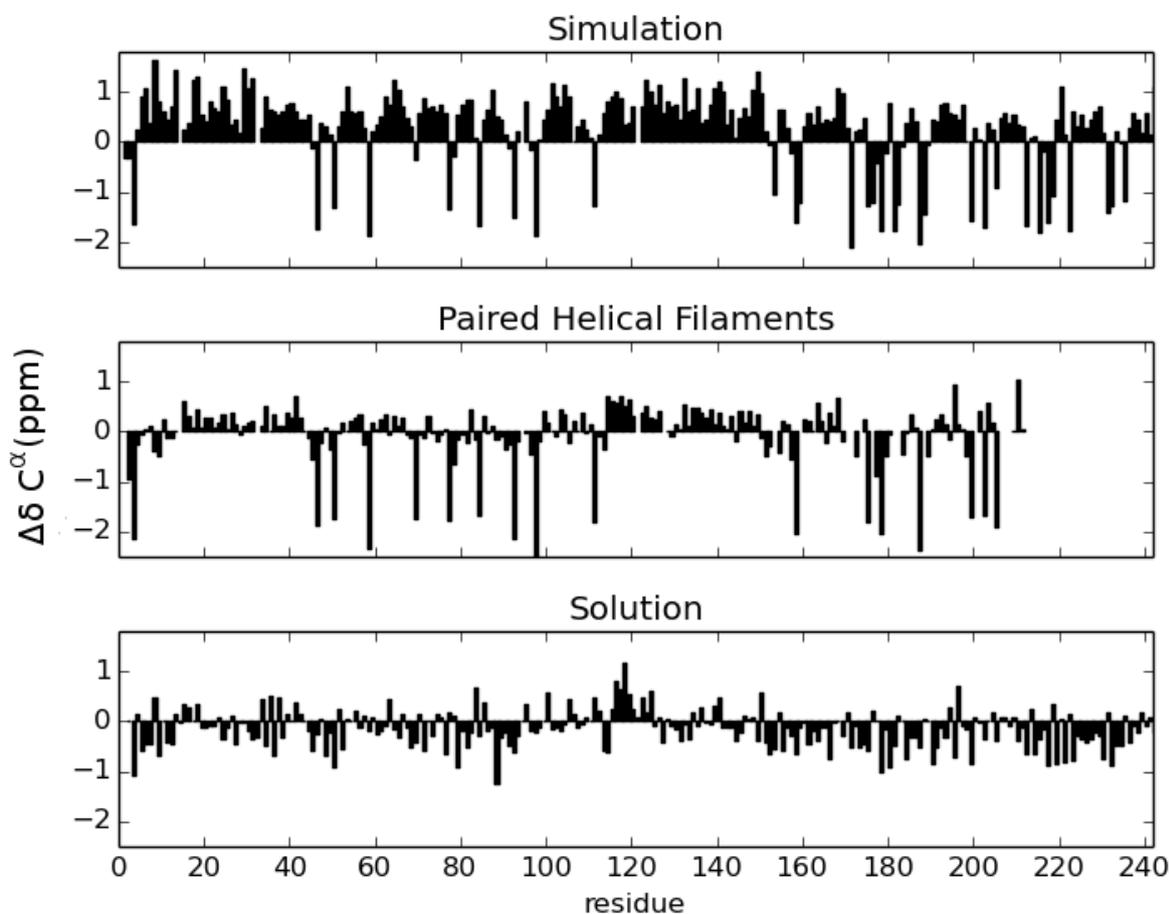

*Figure 4 Chemical Shifts of Simulated TPD compared to PHF and Solution tau.* Our simulations are broadly more similar to PHF tau [23] than to tau in solution [24]. Solution tau NMR was performed over the whole molecule[24].

The chemical shifts (fig. 4) look similar to those found in tau paired helical filaments[22] from the BMRB database[23] rather than tau in solution[24], using chemical shift corrections from[25] as in[24]. Moreover, there are no significant differences in chemical shifts when comparing groups with differing PCA classification. This suggests that the adaptive box method is better at describing contacts that occur in paired helical filaments than it is at showing tau's behavior in solution.

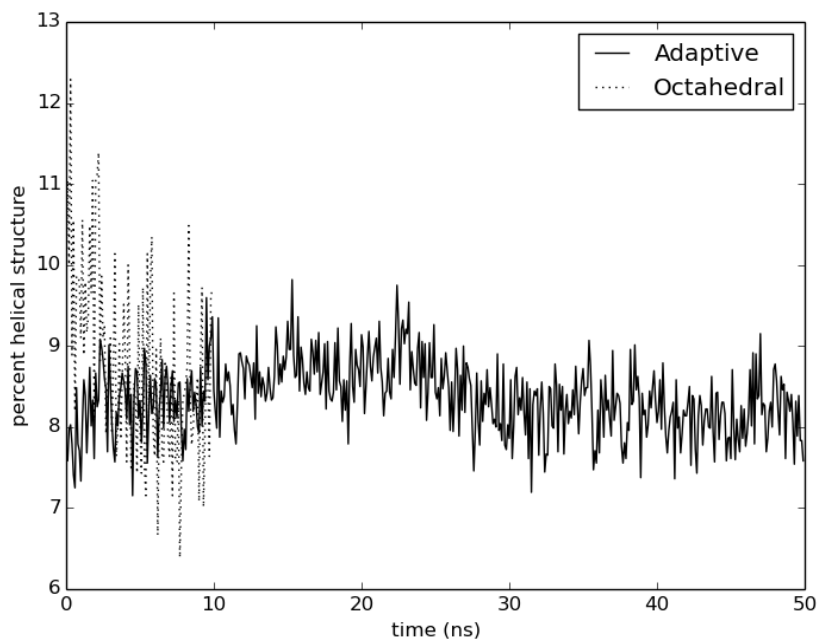

*Figure 5 Percent helical structure over time.* There is more helical structure (alpha and 3-10 helices) in our simulations than should be found in a disordered protein. The octahedral box appears to be converging to a similar distribution as the adaptive box. The secondary structure most likely arises due to using a forcefield not optimized for IDPs.

It appears that the AMBER simulations favor helical secondary structure, though the locations of helices appear to be random. We checked the secondary structure obtained directly via dihedral angles from our PDBs for helical (3-10 or alpha helices), beta and polyproline structures. No beta or polyproline was found of length greater than three residues. In solution NMR of the full-length tau, three 6-8 residue sections of beta structure occurred within the first 242 residues, as well as three small polyproline helices [24]. Beta structure was counted as such when seen 17 percent of the time, so some simulations lacking this structure are not unexpected. One ten-residue alpha helical domain was found in solution NMR [24], although our simulation did not find any large helical domains. While individual simulations may have a few small helical clumps of 3-6 residues persisting for most of the simulation, there is no reproducible secondary structure between these runs, including runs with identical starting conformation but different random number seeds. This is clearly indicated by the overall fractional time spent in helical configurations per residue (fig. 5), and suggests that our simulations not only remain mostly disordered in the ff12SB force-field, but may lack beta and polyproline II content. These secondary structures may not have had enough simulation time to form, and therefore may need to be manually placed in the structure to be simulated accurately.

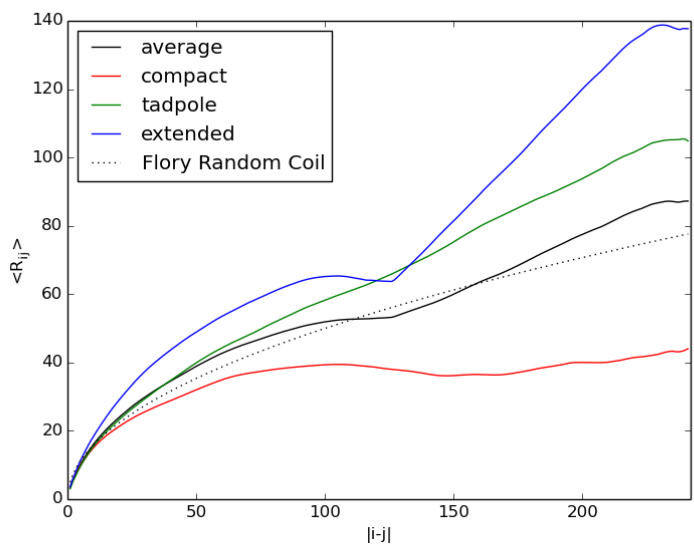

*Figure 6* $R_{ij}$ *profiles of all runs (black), as well as groupings based on PCA categories, compared with expected shape for a Flory random coil.* Compact structures are shown to be globules, while the tadpole and extended structures are not.

### 3.2 Comparison to model classifications for intrinsically disordered proteins

We calculated the $R_{ij}$ profile as in[20] (fig. 6). The TPD appears to be similar to a Flory random coil when looking at an overall ensemble average. However, splitting into the categories defined by PCA changes the shape of the profiles significantly. The compact structures have a profile like that of a globule, while the tadpole and extended structures are more complex. The extended structures, in particular, look more like Flory random coils at lower sequence separations, but are dramatically more extended at higher sequence separations than expected from this type of random coil. These differences suggest that viewing the ensemble overall may appear as a random coil, but our adaptive box model exhibits more complex behavior.

### 3.3 Comparison of Octahedral Box to Adaptive Box

SAXS data extracted from octahedral box runs are very similar to the adaptive box runs. The octahedral box PCA are also comparable with those from adaptive box runs, although the time and extent of each run makes it difficult to compare. Nevertheless, all three types of PCA identified conformations occur if all octahedral box runs are utilized. The ff10 and ff99SB runs extend even further over the 10ns. The ff03r1 and ff03ua runs fall into the extended category, though it is difficult to tell if they would collapse further with more simulation time. The ff12SB run formed a tadpole shape, unlike its adaptive counterpart which remained more extended. Because the octahedral box runs are short (10ns) and are few in number, calculating the chemical shifts is not very informative. However, the secondary structure of the octahedral box runs is comparable to the first 10 ns of adaptive box runs with respect to the percent of time spent in a particular structure (fig. 5). Despite most octahedral box runs beginning with the same conformation, any particular inferred secondary structure is not reproducible for any of the runs. In addition, the force field chosen does not change the lack of non-helical secondary structure. The average helical structure over time appears to be consistent with the longer adaptive box simulations.

*3.4 TIP4PD Water Model comparison*

TIP4PD has been tested and is more accurate than TIP3P at simulating IDPS[13]. We saw a significant difference in behavior depending on the initial state of the TPD. The initially compact run did not extend during the 20ns simulation time, but the initially extended TPD grew longer during a 10ns simulation time. There is also a noticeable difference in the SAXS curves of the extended structure as compared to the compact structure(fig. 7). The extended TIP4PD SAXS curve is nearly representative of the average SAXS curve over all 30 adaptive box runs. There are no significant differences in helical structure occurrence or chemical shifts for the TIP4P-D water model, perhaps due to insufficient simulation time.

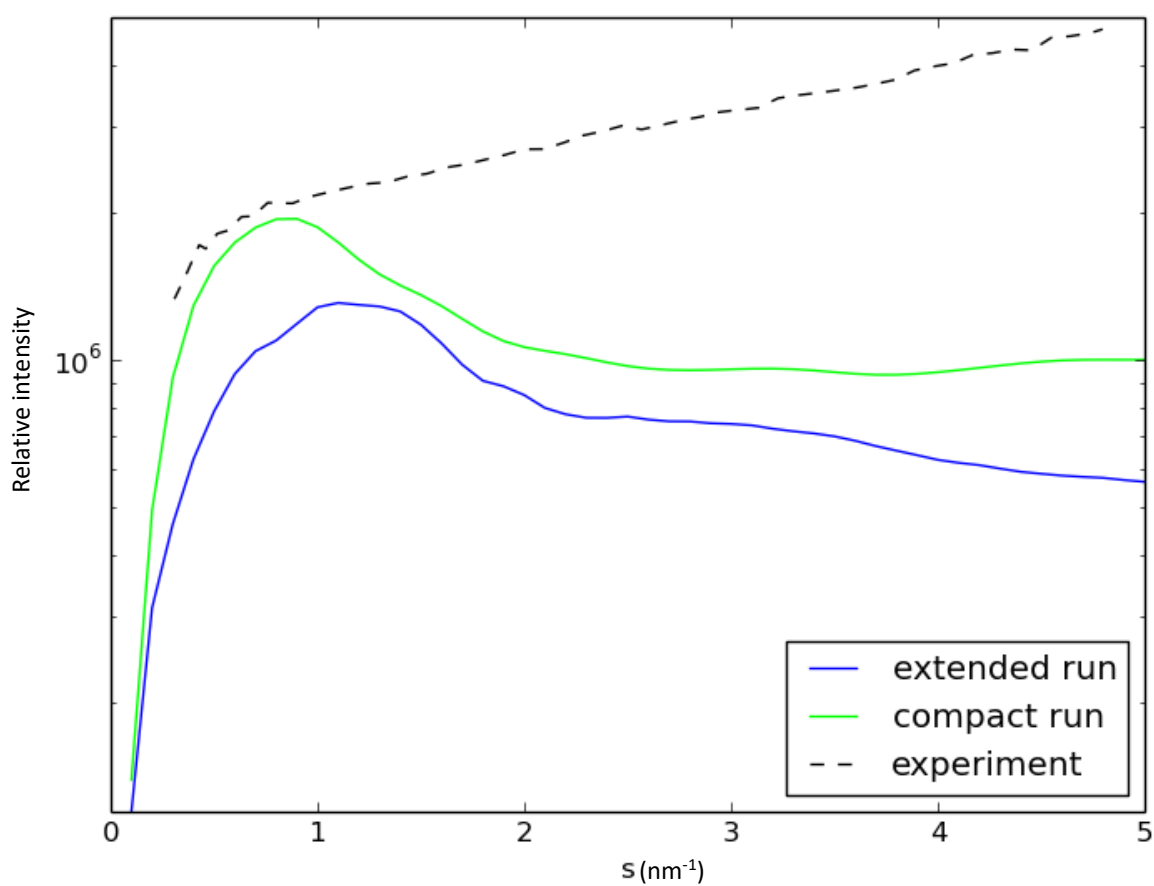

*Figure 7 SAXS plots of TIP4P-D runs compared with K23.* The extended run is very similar to the average of all adaptive box runs. Both TIP4P-D runs exhibit more secondary structure propensities than K23 experiment, but are still mostly unstructured.

## Discussion

We find both the large octahedral box and the adaptive box simulations of TPD fragments are not dimensionally compatible with inter-microtubule spacing in microtubule bundles. In fact, our adaptive box simulations are more similar in size[26] and chemical shift to paired helical filaments than to tau in solution. Because tau in solution has smaller radius of gyration as well, this supports the hypothesis that the observed 20±4 nm length inter-microtubule tau structures must arise from tau complexes, minimally a dimer[27]. On the other hand, the NMR chemical shifts inferred from our simulations are qualitatively similar to the shifts from the projection domain in paired helical filaments rather than for full length tau monomers. This suggests that a study of the projection domain is more relevant for building models of PHFs than the full length tau in solution, which is reasonable given the role the tau binding domain plays in beta sheet formation in the paired helical filaments. In order to simulate a function TPD, we may need to use IDP-specific force-fields that don't create as much helical content, such as CHARMM22*[14].

## Acknowledgements

We acknowledge support from the US NSF via grants DMR-1207624 (N.E.H., D.L.C., R.R.P.S., and N.R.H.) and for computing support US NSF Grant DMR-0844115 (N.E.H., D.L.C., and N.R.H.).

## 4 Author Contributions

Natalie Sanborn performed simulations, analyzed data and had primary responsibility for writing the manuscript. N. Robert Hayre developed the adaptive box algorithm and helped mentor and supervise in early phases of the simulations. Rajiv R.P Singh helped in conceptualizing the model, critically questioning the simulations and analysis, and writing the paper. Daniel Cox is the primary advisor and overseer of the research work and was the secondary writer of the paper.